\documentclass[conference]{IEEEtran}
\usepackage{cite}
\usepackage{amsmath,amssymb,amsfonts}
\usepackage{algorithmic}
\usepackage{graphicx}
\usepackage{textcomp}
\usepackage{xcolor}
\usepackage{url}
\usepackage{verbatim}

\usepackage{scrextend}
\usepackage{subcaption}
\usepackage{listings}

\usepackage[pdftex]{hyperref}
\usepackage{tikz}
\newcommand\copyrighttext{%
    \footnotesize \textcopyright 2026 IEEE. Personal use of this material is permitted.
    Permission from IEEE must be obtained for all other uses, in any current or future
    media, including reprinting/republishing this material for advertising or promotional
    purposes, creating new collective works, for resale or redistribution to servers or
    lists, or reuse of any copyrighted component of this work in other works.
    DOI: \href{https://doi.org/10.1109/ICST69053.2026.00088}{https://doi.org/10.1109/ICST69053.2026.00088}}
\newcommand\copyrightnotice{%
    \begin{tikzpicture}[remember picture,overlay]
        \node[anchor=south,yshift=10pt] at (current page.south) {\fbox{\parbox{\dimexpr\textwidth-\fboxsep-\fboxrule\relax}{\copyrighttext}}};
    \end{tikzpicture}%
}

\def\BibTeX{{\rm B\kern-.05em{\sc i\kern-.025em b}\kern-.08em
    T\kern-.1667em\lower.7ex\hbox{E}\kern-.125emX}}
\begin{document}

\title{Intelligent resource prediction for SAP HANA continuous integration build workloads}

\author{\IEEEauthorblockN{Torsten Mandel}
\IEEEauthorblockA{\textit{SAP SE}\\
Walldorf, Germany \\
torsten.mandel@sap.com}
\and
\IEEEauthorblockN{Jonathan Bader}
\IEEEauthorblockA{Technical University of Berlin \\
Berlin, Germany \\
jonathan.bader@tu-berlin.de}
\and
\IEEEauthorblockN{Hanyoung Yoo}
\IEEEauthorblockA{\textit{SAP Labs Korea} \\
Seoul, Republic of Korea \\
h.yoo@sap.com}
\and
\IEEEauthorblockN{Stephan Kraft}
\IEEEauthorblockA{\textit{SAP SE}\\
Walldorf, Germany \\
stephan.kraft@sap.com}
}

\maketitle
\copyrightnotice

\begin{abstract}

Large enterprises often operate extensive Continuous Integration (CI) pipelines on large, heterogeneous compute clusters, where conservative, statically defined resource requirements are used to ensure build reliability.
This practice leads to substantial system memory over-allocation, reduced cluster utilization, and increased operational costs.

In this paper, we motivate the need for intelligent resource prediction by analyzing over 300,000 historical build executions from a production CI environment with more than one thousand compute nodes.
Our analysis shows that, on average, more than 60\% of allocated system memory remains unused.

We then compare multiple machine learning approaches for predicting build task memory usage, including classification-based methods and regression-based quantile prediction.
Our final solution employs a LightGBM–XGBoost quantile regression ensemble optimized to minimize under-allocation while reducing over-provisioning.
We integrate this solution into the production CI pipeline via a microservice-based orchestration layer, achieving average memory savings of approximately 36GB per build and reducing under-allocation rates to below 0.3\% without negatively impacting build execution times.

\end{abstract}


\section{Introduction} \label{introductionSection}

The business-critical nature and complexity of very large database management systems (DBMS) demand high-quality practices in the software development process. 
In Continuous Integration (CI), automated builds and tests of code changes are frequently triggered to gain quality insights and to detect defects early in the development process. 
The infrastructure for running CI environments of DBMS at scale can incur significant operational costs. 
In the case of SAP HANA~\cite{faerber2011}, the respective CI environment consists of more than a thousand compute nodes.
Due to the size of the application and the complexity of test scenarios, each CI execution run requires hundreds of CPUs and thousands of gigabytes (GB) of system memory.

The source code of SAP HANA consists of more than 40 million lines of code and comprises mainly C++, C, and Python code. The application code is distributed over 350 components representing separate parts of the software project. Source code is managed in a \textit{monorepo}, similar to the approach employed by industry leaders such as Google~\cite{potvin2016}, Microsoft~\cite{harry2017}, Meta~\cite{durham2014}, Uber~\cite{lucido2017}, and BMW~\cite{schwendner2025}.

On average the repository is modified by 800 daily commits triggering automated builds and tests. 
C++ build jobs are executed on a custom distributed build cluster. 
For every commit the builds have to be performed for multiple platforms, e.g., x86\_64, ppc64, aarch64, and build types, e.g., optimized, release, debug. 
While the distribution of build jobs across the cluster significantly reduces the build times, it induces high scalability demands on the cluster management system~\cite{hindman2011} and the availability of cluster resources.

Each executable task on the cluster has a defined set of operational requirements such as CPU and main memory. The cluster management system matches these task requirements to incoming resource offers. Large resource requirements typically lead to longer wait times for a suitable resource match. In addition, operational costs in modern cloud environments are often based on pay-per-use contracts, resulting in higher costs for tasks with high resource demands.

The resource requirement prediction of build jobs is challenging as they depend on a number of dynamic and interdependent factors like code change complexity, development branch, compiler parameters, compiler optimizations, and target platform. 
In practice the definitions of system memory requirements are often static and based on expert knowledge and heuristics.
Our goal is to move assigned memory closer to each job's actual requirement while keeping build performance stable and avoiding under-allocations that lead to costly out-of-memory (OOM) events.
Our main contributions are as follows.

\begin{itemize}
    \item We provide a large-scale characterization of memory allocation inefficiency in a production SAP HANA CI environment. To this end, we analyze more than 300{,}000 build job executions on a heterogeneous cluster with over 1{,}000 compute nodes and release the underlying dataset publicly~\cite{SAPDataSet}.

    \item We evaluate multiple machine learning approaches for CI build task memory prediction and show that a conservative LightGBM-XGBoost quantile ensemble provides the best trade-off between memory savings and reliability on the offline data.

    \item We deploy the selected prediction approach in a production CI pipeline using a microservice-based orchestration architecture and report average memory savings of approximately 36\,GB per build without negatively impacting build execution times.
\end{itemize}

\section{Problem Statement} \label{problemStatementSection}

\begin{figure}[t]
        \centering
        \includegraphics[width=\linewidth]{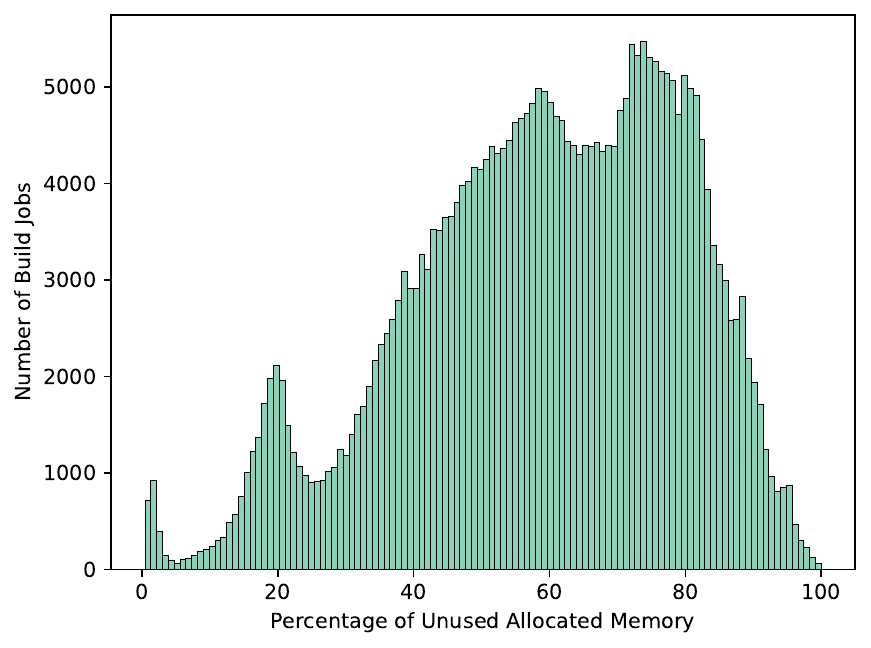}
    \caption{Percentage of Unused Memory over all successfully executed tasks.}
    \label{fig:sap-dataset-overview}
\end{figure}

To better understand the memory allocation behavior in the SAP HANA CI build clusters, we analyzed the memory usage of more than 300,000 build jobs over an eight-week period.
Our data show that approximately 8.6\% of all jobs failed due to memory under-allocation, resulting in costly job restarts.
While the remaining 91.4\% of jobs did not fail due to memory under-allocation, the amount of reserved but unused memory is substantial.
Figure~\ref{fig:sap-dataset-overview} shows the distribution of the percentage of unused but allocated memory, with a median of 60.9\% unused memory.
In absolute terms, this corresponds to 33.71~petabytes of unused main memory over the eight-week period.

A closer look at the data and the baseline allocation strategy reveals that memory is not assigned continuously, but selected from a fixed set of only 22 predefined memory bins referring to the respective underlying job.
As a result, builds with significantly different actual memory requirements are often mapped to the same coarse-grained allocation class.

This systematic over-allocation directly impacts the scalability of the CI infrastructure: fewer build jobs can be scheduled in parallel, leading to increased queueing times and delayed developer feedback.
To address these limitations, we collaborated with the Technical University of Berlin to investigate data-driven methods for predicting the memory requirements of build jobs.
The results of this collaboration have been presented in prior work~\cite{bader2025memopt} and serve as the motivation and foundation for the implementation described in this paper.

During this collaboration we built a pipeline to tune hyperparameters and evaluate various memory prediction methods on the collected offline dataset.
We first explored classification methods for predicting peak main memory consumption.
Our idea was to dynamically assign a memory bin to a given job instead of static assignment.
After prediction, we added an offset policy that assigned the next larger bin to a given job to avoid failures due to under-provisioning. 
We evaluated LightGBM, XGBoost, Random Forest and Logistic Regression.

\begin{figure*}[t]
        \centering
        \includegraphics[width=0.7\linewidth]{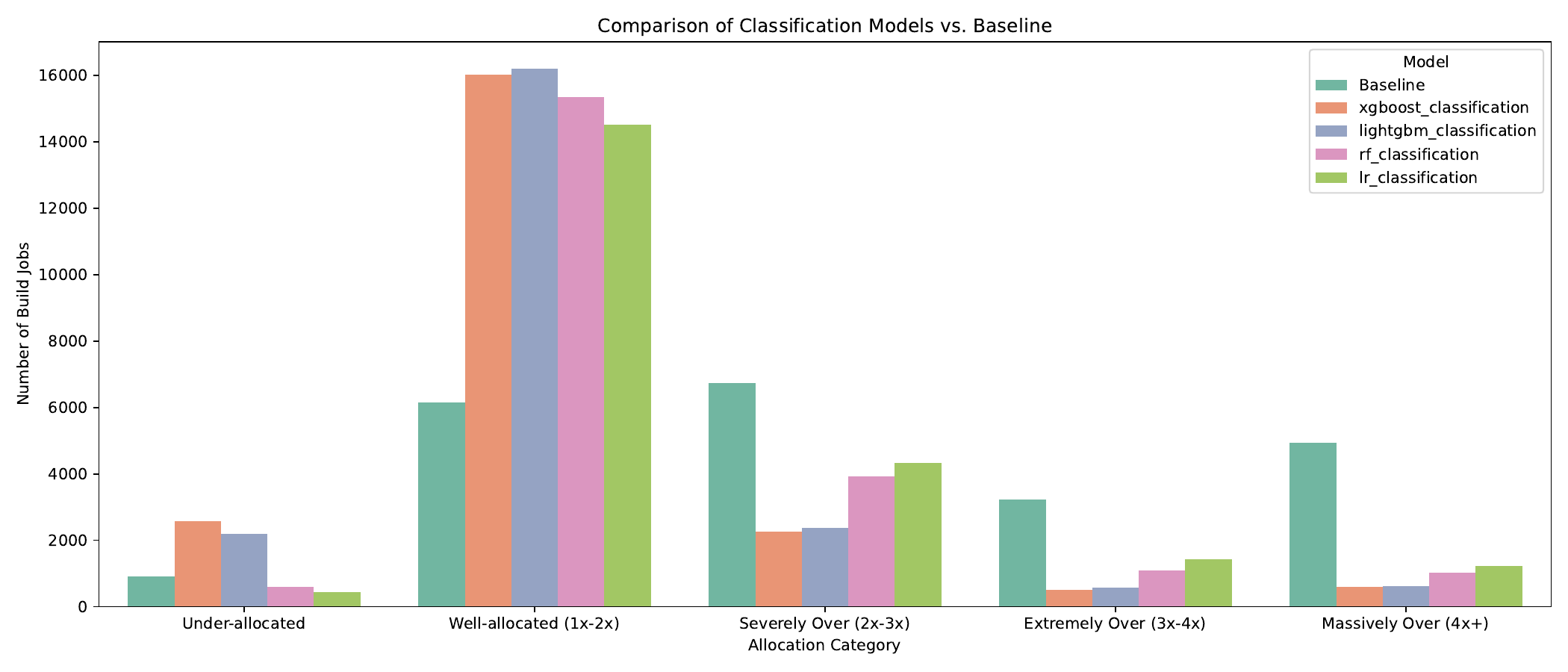}
    \caption{Comparing the memory prediction of various classification methods with the baseline memory allocation.}
    \label{fig:sap-results-classification}
\end{figure*}

Figure~\ref{fig:sap-results-classification} compares the memory allocation quality of the four classification approaches with the baseline strategy.
The x-axis represents the allocation quality, e.g., well-allocated build jobs reserve between 100\% and 200\% of their actual peak memory consumption, while under-allocated jobs correspond to build failures caused by out-of-memory events.
The baseline exhibits a large fraction of over-allocated jobs (2×–4×+), whereas the proportion of well-allocated jobs remains comparatively low.
In contrast, all classification methods substantially increase the share of well-allocated jobs. However, the most conservative classifiers, which achieve the lowest under-allocation rates, also exhibit higher degrees of over-allocation compared to less conservative alternatives.

These results indicate that a conservative prediction strategy is required to reliably avoid under-allocation.
Consequently, we investigated regression-based approaches that directly predict upper memory quantiles (e.g., P95) instead of mean values.
We evaluated single quantile regressors based on LightGBM and XGBoost, as well as homogeneous and heterogeneous ensemble combinations involving LightGBM, XGBoost, and CatBoost, each optimized through an independent hyperparameter search.

\begin{figure*}[t]
        \centering
        \includegraphics[width=0.7\linewidth]{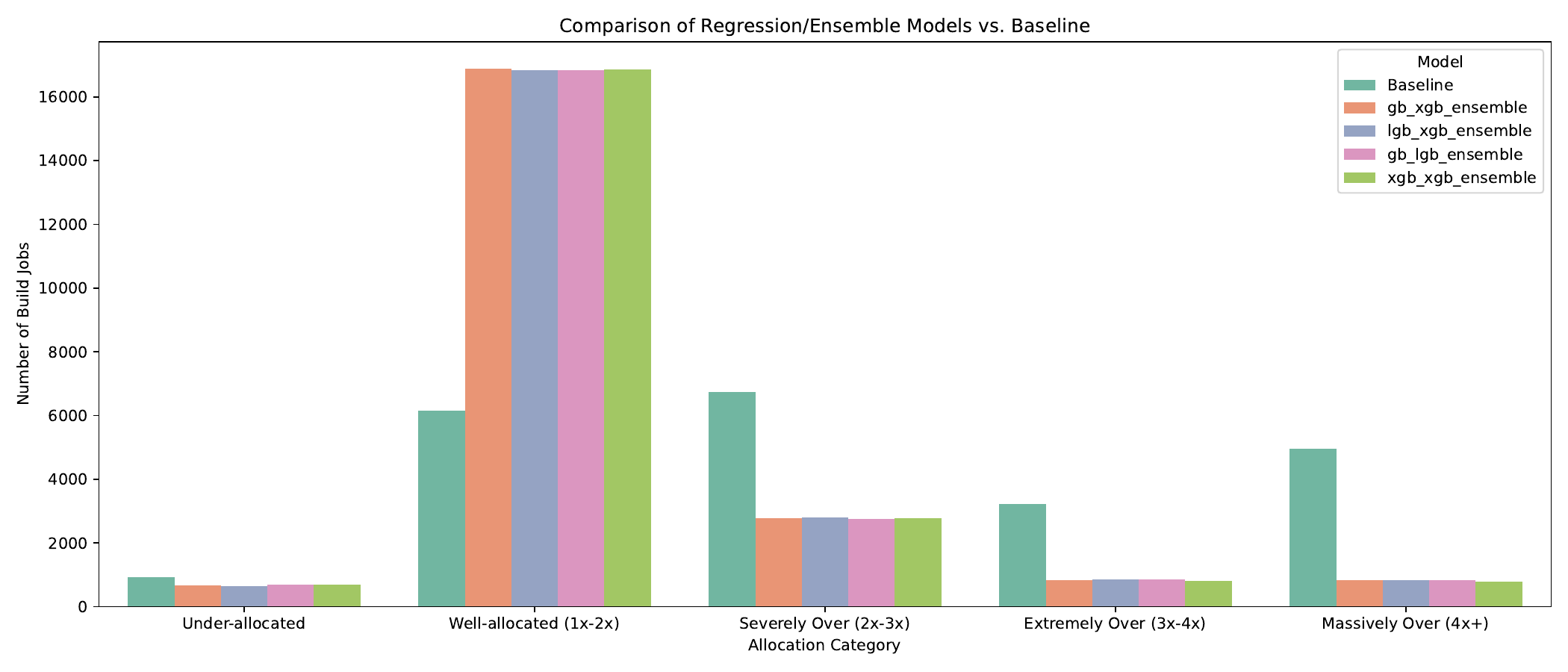}
    \caption{Comparing the memory prediction of various (ensemble) regression methods with the baseline memory allocation.}
    \label{fig:sap-results-regression}
\end{figure*}

Figure~\ref{fig:sap-results-regression} compares the four best-performing regression setups with the baseline allocation strategy.
The baseline performs worst, exhibiting both the highest under-allocation rate and the lowest proportion of well-allocated build jobs.
In comparison, the evaluated regression methods outperform the baselines but all show a very similar performance.
Among them, the LightGBM–XGBoost ensemble achieves the best overall results, yielding the lowest number of under-allocations while maintaining a high proportion of well-allocated jobs.

While these offline experiments on an initial dataset demonstrated that data-driven memory prediction can substantially improve allocation quality, translating these results into a production CI environment poses additional challenges.
In the following, we describe step by step how we transferred the insights gained from our collaborative evaluation into a production-ready memory prediction service deployed in the SAP HANA CI infrastructure.


\section{Prediction model development} \label{experimentSection}
In this section, we describe the model development process for predicting the memory requirements of CI build jobs. 
We first introduce the dataset used for offline modeling, followed by the methodology and the classification-based and regression-based approaches evaluated in this work.

\subsection{Dataset} \label{dataSection}
Automated builds are triggered for every submission to the central code repository. 
For each build operation its inherent build parameters and observability data like CPU, system memory, disk, and network usage are persisted in a central database. 
These data enable the development of offline memory prediction models that can be evaluated without impacting ongoing CI operations, while providing insight into the potential for improved build task memory prediction.

As only a subset of data fields is related to the memory requirements of build jobs, we reduce the dataset prior to model training. 
The cluster management system of the CI infrastructure does not allow the dynamic interpolation of resource requirements once the execution of a task has started. 
Therefore the data fields represented by continuous timeseries, e.g., memory measurements in one second intervals, are pre-aggregated and represented by their maximum values. 
This approach reduces the exported dataset by a factor of $160 \pm 222$. 
Note that related work has shown that training a memory timeline prediction on the full and un-aggregated dataset can yield significant benefits, as the final linking stage of a build job often has plateaus of increased memory usage~\cite{bader2024ks+}. 
However, since dynamic interpolation of resource limits is not supported by the cluster management system, we restrict our modeling to pre-aggregated peak values.

The historical trend of the pre-aggregated data extract is persisted in a Git repository, as the retention policy frequently moves the high-resolution observability data out of the central database. 
Typical data exports are in the range of 10k-15k builds per week and we concatenate up to 16 weekly exports\footnote{This choice trades off additional training data against repository evolution, where stale training data could worsen prediction quality. One option to balance both effects would be to associate a training weight with each sample according to its age.}. 
A sample dataset export is publicly available on GitHub~\cite{SAPDataSet}.

The sample dataset comprises 22 memory allocation classes depending mainly on architecture, location, and build profile settings. 
The statistical distribution across those classes is such that 6 classes already cover 80\% of the parameter input space and 10 classes cover 90\%.
Many of those parameter tuples are assigned to a memory class that does not fit their actual needs, e.g., because the empirical assignment does not consider a crucial detail in one of the exported parameters. A machine learning (ML) model can learn these dependencies and assign a better-fitting memory class.

\subsection{Methodology}
The initial exploration of data and models for the classification approach (\ref{LinearSVCSection}) was done in a Jupyter~\cite{Jupyter} notebook project and later expanded into an ML pipeline similar to the one described in~\cite{bader2025memopt}. 
The main steps of the respective pipelines can be summarized as follows:
\begin{itemize}
    \item Data extraction
    \item Feature engineering\footnote{\label{AppliesOnlyTo}Only for \ref{LGB_XGB_Section}}
    \item Splitting into training and test datasets
    \item Hyperparameter optimization\footref{AppliesOnlyTo}
    \item Model training
    \item Evaluation and visualization
\end{itemize}
The development was primarily performed in Python using scikit-learn~\cite{SKLearn} and related machine learning libraries~\cite{LightGBM,XGBoost}.
Data pre-processing is performed with Pandas~\cite{Pandas}. The code of the models used in Section~\ref{LGB_XGB_Section} is publicly available on GitHub~\cite{SAPResourceOptimizer}.

\subsection{Classification-Based Memory Prediction} \label{LinearSVCSection}
In the initial exploration, we investigated LinearSVC from scikit-learn~\cite{SKLearn} and classified builds into two categories that either fell below or above a memory requirement threshold.
The model evaluates to \textit{true} if the predicted memory requirement lies below the defined threshold for a given input of features.
Listing~\ref{listingClassification} formalizes how we use these predictions to determine the memory requirement for a given build. If the model predicts that the build falls below the threshold, the refined memory requirement is set to the safeguarded threshold. The \texttt{min} operation ensures that the classifier can only reduce the baseline configuration and never increase it. During our experiments, a safety factor of two yielded good results.   

\begin{lstlisting}[language=Python, breaklines, caption=Pseudocode Classification Approach, label={listingClassification}, frame=single, basicstyle=\footnotesize\ttfamily]
# featureVector: feature vector of a build
# threshold: defined threshold for the build in GB
# safetyFactor: a multiplicative safety factor
# originalConfiguration: the original empirical memory configuration

belowThreshold: bool = predict(featureVector, threshold)
memory = originalConfiguration
if belowThreshold:
  memory = min(threshold * safetyFactor, memory)
return memory
\end{lstlisting}

For training and testing we selected the features presented in Table \ref{UsedFeaturesTable} from the exported dataset (\ref{dataSection}). 
Features are encoded with a DictVectorizer\cite{SKLearn} for feature vectorization.

We used max\_rss\_gb (the maximum resident set size during the build, in gigabytes) as the optimization target, accepted an initial false-positive rate of 10\%, and set the threshold to 50GB. 
As mentioned above, we apply a safety factor of two for two reasons. First, it reduces the effective false-positive error that could lead to an OOM event to less than 0.2\%. Second, we need some additional free system memory for the page cache so that turnaround times are not negatively affected.

\subsection{Regression-Based Memory Prediction} \label{LGB_XGB_Section}
In the second evaluation step, we investigated multiple regression models and model ensembles on the offline dataset described in~\cite{bader2025memopt, SAPResourceOptimizer}. 
Several of the evaluated models performed better than the initial classification model from Section~\ref{LinearSVCSection}. In this section, we focus on the best ensemble, which combines a LightGBM~\cite{LightGBM} regression model and an XGBoost~\cite{XGBoost} regression model and outputs the maximum prediction of the two submodels. We highlight three parts of the pipeline: feature engineering, hyperparameter search, and final training with the selected hyperparameters.

\subsubsection{Feature Engineering}
To increase prediction quality, we engineer synthetic columns that decouple information which is more informative when analyzed separately, e.g., temporal decomposition (year, month, day of week, etc.) and rolling-window features that capture temporal correlations of simultaneously scheduled builds.
In particular, the build profile can often be decomposed into compiler, CPU architecture, and optimization level to analyze the impact of each separately.
The full list of used features and their explanations in our re-implementation is given in Tables~\ref{UsedFeaturesTable} and~\ref{FeaturesTable}.

\subsubsection{Hyperparameter Search}
Both LightGBM and XGBoost offer several tunable initialization parameters affecting the fitting functions. For optimization of the best prediction results towards
a given scoring function a hyperparameter optimization for the conjoined models using Optuna optimizer\cite{optuna} is being performed.
The scoring function for a given hyperparameter set $\theta$ is given by \cite{SAPResourceOptimizer}:
\begin{equation}
    \underset{\theta}{min} Cost(\theta) = c * \sum under\_alloc(\theta) + \frac{\sum{pred\_rss(\theta)}}{\sum actual\_max\_rss}
\end{equation}
where we configured $c=3$ for a more aggressive optimization compared to the value of $c=5$ given in \cite{bader2025memopt}. The summation runs over each prediction / actual value of the training set for a given hyperparameter trial.

\subsubsection{Training}
After a configured number of hyperparameter optimization steps has been completed, we train the ensemble once more using the best parameter set and persist the resulting model and its metadata. In contrast to the classification approach from Section~\ref{LinearSVCSection}, the regression model predicts a numerical memory value in GB for a given build parameter set. For our infrastructure, we made three practical adjustments to the original model implementation~\cite{SAPResourceOptimizer}. First, we apply the safety factor only for the baseline, final evaluation, and production inference, but not during the Optuna scoring loop, because this improved convergence of the hyperparameters. Second, if the safeguarded prediction exceeds the original empirical requirement, we keep the original requirement unchanged. This is the deployed policy and ensures that the regression-based refinement can reduce memory assignments but does not increase them. Third, we use a safety factor of 1.2 instead of the factor of 2 used in Section~\ref{LinearSVCSection}. The target value is again max\_rss\_gb and the used feature names are displayed in Table~\ref{UsedFeaturesTable}.

\begin{table}[htbp]
\caption{Used features}
\begin{center}
\begin{tabular}{|c|c|}
\hline
\textbf{LinearSVC}&\textbf{LGB-XGB-Quantile-Ensemble} \\
\hline
branch & branch \\
buildProfile &  \\
jobs & jobs \\
location & location \\
makeType & makeType \\
targets &  \\
localJobs & localJobs\\
component & component\\
max\_rss\_gb & max\_rss\_gb \\
 & bp\_arch \\
 & bp\_compiler \\
 & bp\_opt \\
 & ts\_year \\
 & ts\_month \\
 & ts\_dow \\
 & ts\_hour \\
 & ts\_weekofyear \\
 & target\_cnt \\
 & target\_has\_dist \\
 & lag\_1\_grouped \\
 & lag\_max\_rss\_global\_w5 \\
 & rolling\_p95\_rss\_g1\_w5 \\
\hline
\end{tabular}
\label{UsedFeaturesTable}
\end{center}
\end{table}

\begin{table*}[htbp]
\caption{Features}
\begin{center}
\begin{tabular}{|c|c|}
\hline
\textbf{Feature name}&\textbf{Explanation} \\
\hline
time & Build execution timestamp \\
branch & (Git) branch of the change being built \\
buildProfile & CMake profile of the change being built (contains information about target architecture, optimization level, compiler etc. ) \\
jobs & Parallelization level of build jobs \\
location & Abstract definition of cloud provider and build cluster specification \\
makeType & Explicit optimization level (sometimes redundant regarding buildProfile) \\
targets & CMake build target (e.g., all, dist) \\
localJobs & Parallelization level on the primary building node \\
component & Project being built \\
max\_rss & Maximum resident set size ever measured during the build in bytes \\
memory\_fail\_count & Kernel memory allocation fail count on the node during execution of the build \\
maxMemoryUsageBytes & Highest ever measured total memory usage in the container during the build in bytes \\
\hline
\textbf{Engineered feature name}&\textbf{Explanation} \\
\hline
bp\_arch & Target architecture (derived from buildProfile if possible) \\
bp\_compiler & Used compiler (derived from buildProfile if possible) \\
bp\_opt & Optimization level (derived from buildProfile if possible) \\
ts\_year & Year of "time" timestamp \\
ts\_month & Month of "time" timestamp \\
ts\_dow & Day of week of "time" timestamp \\
ts\_hour & Hour of "time" timestamp \\
ts\_weekofyear & Week of year of "time" timestamp \\
target\_cnt & Count of targets (splitting targets string by ",") \\
target\_has\_dist & Flag on the occurrence of the "dist" target in the target list\\
lag\_1\_grouped & Group by used data columns key and shift by 1\\
lag\_max\_rss\_global\_w5 & Rolling mean window over used feature grouping with 5 entries \\
rolling\_p95\_rss\_g1\_w5 & Rolling mean window over used feature grouping with 5 entries, 95\% quantile \\
max\_rss\_gb & Max RSS scaled to gigabytes, target of the optimization \\
\hline
\end{tabular}
\label{FeaturesTable}
\end{center}
\end{table*}

\subsubsection{Feature Term Importance}
\begin{figure*}[htbp]
\centerline{\includegraphics[scale=.6]{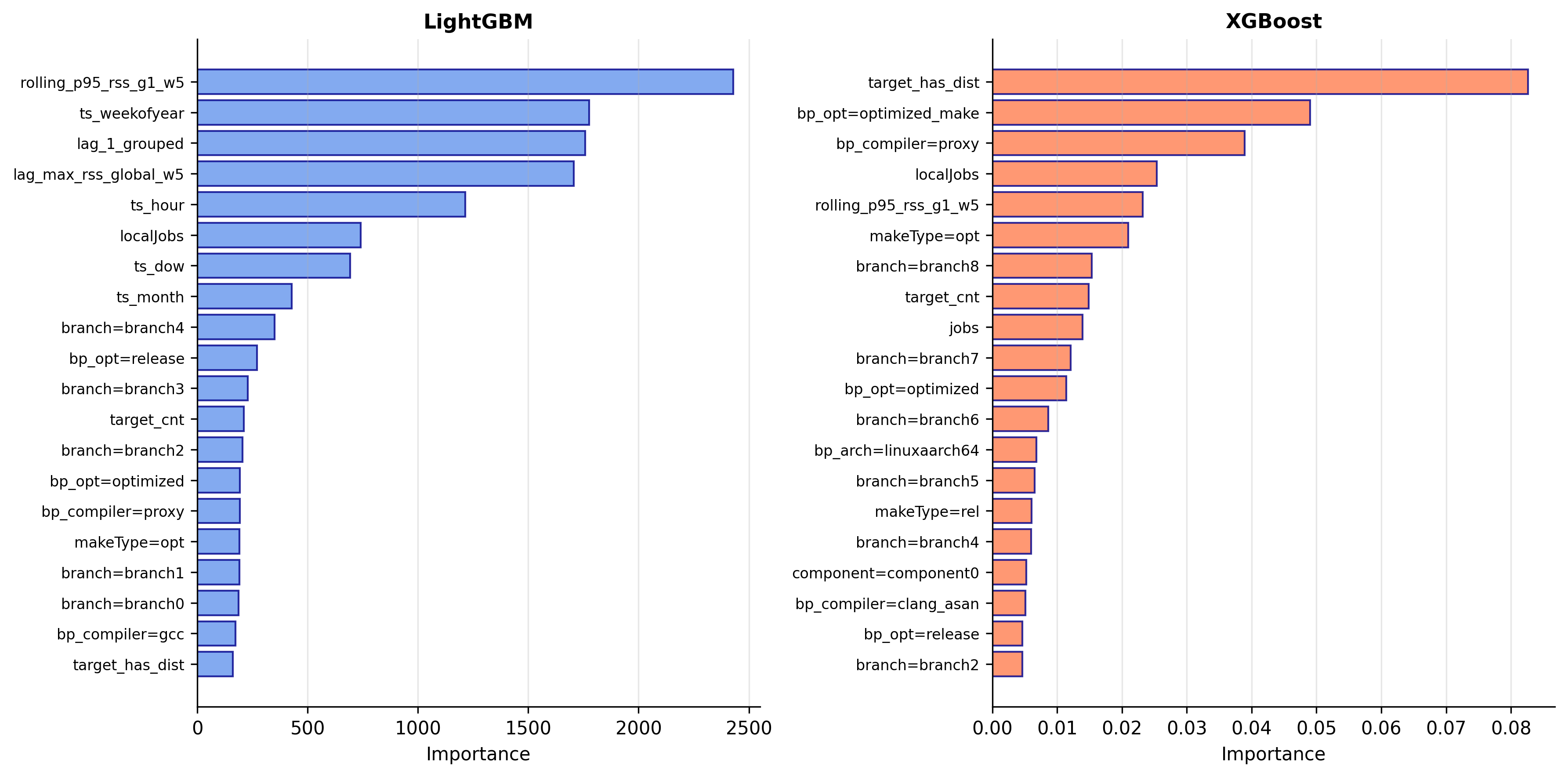}}
\caption{Feature term importance of the respective models of the LGB-XGB-Ensemble, x axis according to each model's internal scaling.}
\label{FeatureTerm}
\end{figure*}

Figure~\ref{FeatureTerm} shows the top 20 feature terms of each model by importance. The most prominent signals are rolling-window features of related builds, execution-time related features, branch names, the number of jobs, and the number of targets. The engineered features bp\_opt and bp\_compiler are more informative than the original buildProfile field, which combines several of those dimensions. In contrast, location contributes comparatively little, indicating that the workload remains sufficiently homogeneous across the supported locations.

\section{Modeling Results} \label{resultsSection}
We now compare the memory savings of the classification and regression methods on the held-out test data. 
The experimental results are expected to reflect the production application closely, except for a limited temporal drift caused by the continuous evolution of the code repository and the corresponding build process.

\subsection{Classification}
\begin{figure}[htbp]
\centerline{\includegraphics[scale=.6]{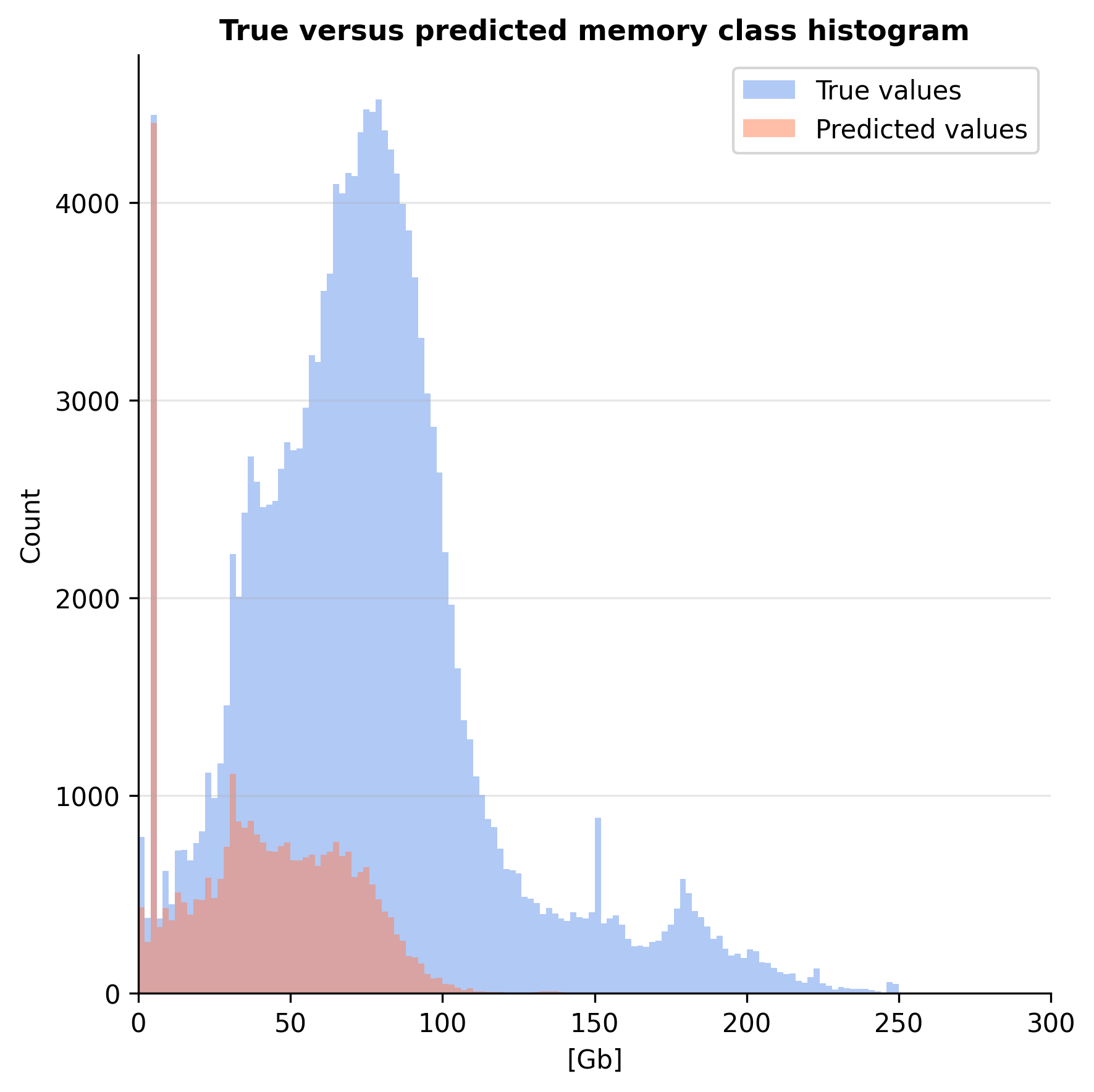}}
\caption{Histogram of actual vs predicted build memory usage.}
\label{Histogram}
\end{figure}

Using the classification model, we observed average memory savings of 9-10GB per build, a further 48GB of theoretical optimization potential per build, and an under-allocation\footnote{Potentially slowing down the build and/or leading to an OOM event.} rate of $0.14\%$ when including the safety factor. 
Figure~\ref{Histogram} compares the actual and predicted memory class usage on the test dataset for the 50GB threshold and safety factor of two. The remaining mismatch, especially in the $\leq$100GB range, indicates that substantial optimization potential is still left unused.

The most influential features are related to non-standard build profiles and branches that usually build only subsets of the full project.

\subsection{Regression}
For the regression model, we evaluate the performance of the models using five memory classes~\cite{bader2025memopt}:
\begin{itemize}
    \item Massively overallocating \\ ($max\_rss_{set}>4 * max\_rss_{req}$)
    \item Extremely overallocating \\ ($3*max\_rss_{req}< max\_rss_{set}\leq 4*max\_rss_{req}$)
    \item Severely overallocating \\ ($2*max\_rss_{req}< max\_rss_{set}\leq 3*max\_rss_{req}$) 
    \item Well allocating \\ ($max\_rss_{req}\leq max\_rss_{set}\leq 2*max\_rss_{req}$) 
    \item Under allocating \\ ($max\_rss_{set} < 1*max\_rss_{req}$) 
\end{itemize}
where "set" refers to the configured memory requirement of a build job (either empirical or model-based) and "req" refers to the actually measured requirement.
The population shift across these categories is shown in Table~\ref{AllocationCategoryImprovement}.

\begin{table}[htbp]
\caption{Population share by memory allocation class}
\begin{center}
\begin{tabular}{|c|c|c|}
\hline
    & Dataset & LGB-XGB- \\
    & baseline & Quantile-Ensemble \\
    \hline
    Massively overallocating & 23.5\% & 3.8\% \\ 
    Extremely overallocating & 14.0\% & 4.01\% \\
    Severely overallocating & 26.5\% & 12.7\% \\
    Well allocating & 28.1\% & 76.6\% \\ 
    Under allocating & 7.9\% & 2.89\% \\
    \hline         
\end{tabular}
\label{AllocationCategoryImprovement}
\end{center}
\end{table}

For direct comparison with the classification model, we define savings per build as the reduction in assigned memory relative to the empirically configured baseline. Under this definition, the regression model achieves mean savings of 37GB per build, with a theoretical further optimization potential of 51GB per build on average\footnote{This remaining potential stems from cases below the classification threshold that the classification model cannot further optimize, as well as from discrepancies that cross the threshold boundary.}, and an average under-allocation rate of $0.28\%$. The comparable baseline under-allocation rate on this dataset is $3.84\%$\footnote{For the regression evaluation, the baseline uses detected kernel allocation failures as a mask for under-allocations. These failures do not necessarily originate from the build itself, because other containers may run in parallel on the same node. Consequently, this baseline may overestimate the number of build-specific OOM events compared to Section~\ref{LinearSVCSection}.}. The standard deviation of the savings is on the order of $\sigma \approx 40$GB, which can be attributed to clusters of empirically highly over-provisioned jobs that skew the distribution upward.

\begin{figure}[htbp]
\centerline{\includegraphics[scale=.6]{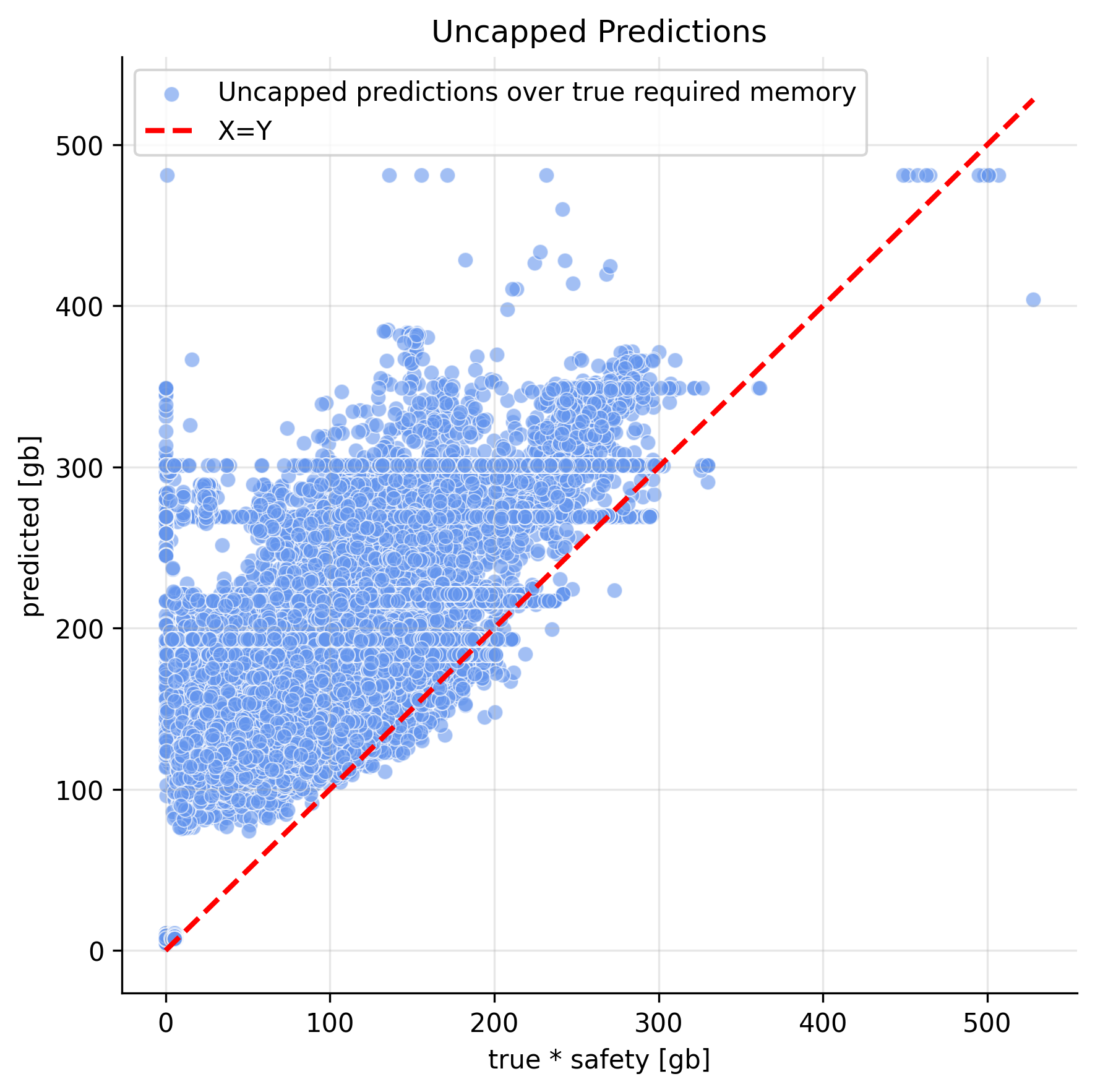}}
\caption{Measured peak memory versus raw regression predictions of the LGB-XGB-Quantile-Ensemble.}
\label{Unclipped}
\end{figure}

\begin{figure}[htbp]
\centerline{\includegraphics[scale=.6]{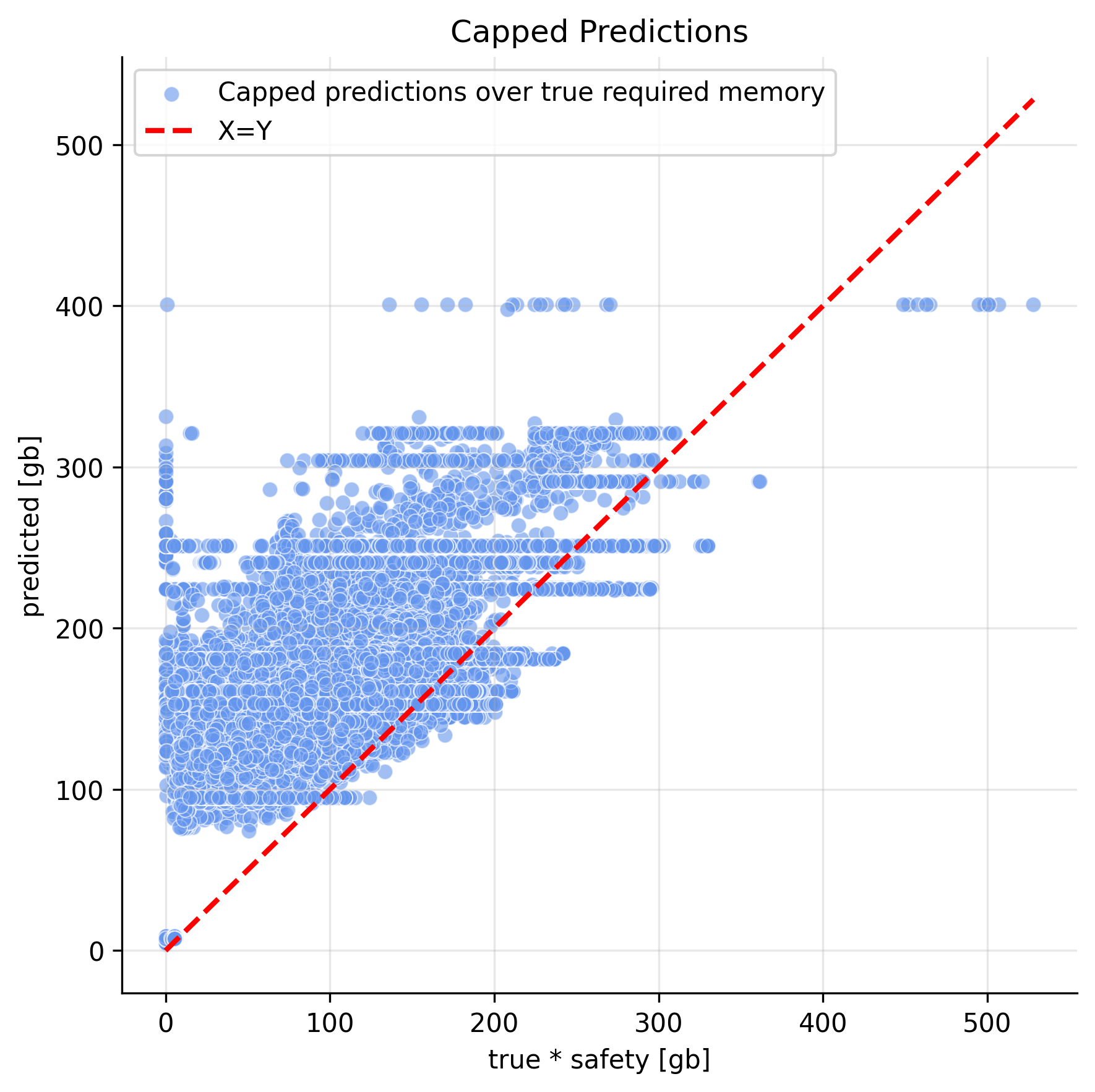}}
\caption{Measured peak memory versus production-clipped regression predictions of the LGB-XGB-Quantile-Ensemble.}
\label{Clipped}
\end{figure}

Figures~\ref{Unclipped} and~\ref{Clipped} compare measured max RSS values with the regression predictions before and after the deployment post-processing. Figure~\ref{Unclipped} shows the raw model output, whereas Figure~\ref{Clipped} applies the production rule that clips refined allocations at the empirical baseline. The red line denotes perfect predictions of true max\_rss memory usage after applying the safety factor. Points below the line indicate potential page-cache pressure or even OOM events, whereas points above the line indicate over-allocation. The similarity of the two charts shows that only a small share of predictions would exceed the empirical baseline; clipping therefore mainly acts as a reliability safeguard. The production deployment uses the clipped variant. In future work, we would like to allow exceeding the empirical setting, but only once we have a more precise mapping of builds that caused their own OOM events rather than relying on the kernel allocation failure count of a node.

Due to the superior performance of the regression model, we selected it for production assessment and deployment. 

\section{Production Application} \label{productiveApplication}

\begin{figure}[htbp]
    \centering
    \includegraphics[scale=.45]{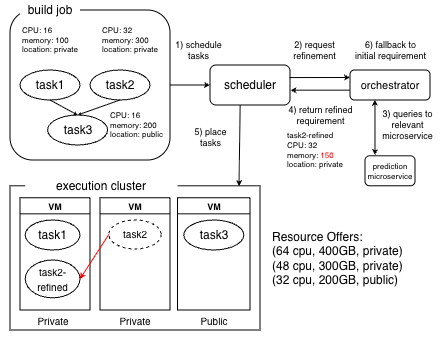}
    \caption{Runtime refinement pipeline for task scheduling and execution in the cluster management system}
    \label{cluster_management_architecture}
\end{figure}

In this section, we present the application of the previously introduced prediction models in the production execution environment of the SAP HANA CI infrastructure. C++ build jobs are executed on a cluster management system that governs distributed compute resources. The compute resources reside in multiple private and public cloud environments employing bare-metal servers and virtual machines. The workloads are defined in a domain-specific language expressed as directed acyclic graphs, where vertices represent executable tasks and edges define data flows and dependencies such as files, directories, or numeric values (see Figure~\ref{cluster_management_architecture}).

The build execution involves complex compile operations and data transfers across multiple platforms and cluster regions. To manage these distributed workloads, the cluster management system uses Apache Mesos~\cite{hindman2011} for resource management. Each task is annotated with tuples of computational resource requirements such as CPU and system memory. In addition, the execution cluster sends tuples of available resource offers to the cluster management system. The scheduler component then matches resource requirements and offers according to its optimization targets. Build jobs are scheduled in containers to enforce resource limits and provide performance isolation.

In the baseline configuration, the resource requirements of build jobs are statically defined based on domain expert experience and heuristics. Such definitions typically overestimate the actual resource demands to safeguard successful build execution. To apply intelligent resource prediction, we extended the cluster management system with an additional software component. The so-called \textit{orchestrator} follows a FastAPI-based~\cite{FastAPI} microservice paradigm and performs dynamic, on-demand refinements of task resource requirements. 

Figure~\ref{cluster_management_architecture} illustrates the runtime pipeline: (1) The task graph is scheduled to the scheduler (2) the scheduler sends a task refinement request to the orchestrator via HTTP, (3) the orchestrator registers the task and dispatches it to the corresponding handler, and the handler queries the relevant prediction microservice, (4) the orchestrator aggregates the responses and returns refined requirements, (5) the scheduler places the task using the refined values, and (6) restart logic handles the rare OOM failures. If a refined build job fails due to an OOM event, the system performs two restart attempts. The first restart keeps the refined memory requirement, while the final restart falls back to the initial unrefined configuration.

The proposed architecture is generic and supports the incremental addition of arbitrary domain-specific prediction services. For the integration of the build memory prediction model, we implemented a dedicated microservice that is queried via REST API. The microservice can use both the classification-based and regression-based memory prediction models.

OOM events of build jobs are logged in a central database, and we plan to use these data points to augment the training data with synthetically increased memory requirements for builds that failed due to OOM.

\subsection{ML Pipeline} \label{SapMlPipeline}
The code base and CI environment evolve continuously, with hundreds of daily commits, regular compiler upgrades, newly added build profiles, and hardware upgrades. Since the prediction model has to reflect these changes continuously, we set up a GitHub Actions~\cite{GithubActions} pipeline to adapt the model in a well-defined process.  

The pipeline includes data extraction, model training, model storage, and evaluation. Unit and integration tests are executed for pull requests and release builds. We implemented dedicated Actions for pull requests that remove temporary test models after evaluation, as well as production update actions triggered by code merges or by a cron schedule. The GitHub Actions workflows check out both the model/service repository and a configuration/data repository, which stores the pre-aggregated infrastructure data in CSV files using Git LFS.

The general workflow for release builds and cron runs can be summarized as follows:
\begin{itemize}
    \item Build a new image if the Dockerfile has changed since the last commit
    \item Instantiate a container with the most recent head image
    \item Run a script to extract data from the infrastructure databases
    \item Pre-split the extracted data into disjoint 60/40\% training and testing CSV files for comparability across trained models
    \item Push the new CSV files to the data repository
    \item Start the script that trains all models configured in the data repository
    \item Push the new models, parameters, and plots to an MLflow~\cite{mlflow} instance
    \item Persist the training configuration and result metadata in the data repository
\end{itemize}

The training artifacts can be evaluated via MLflow or via dedicated Grafana~\cite{Grafana} dashboards.

\subsection{Measurement results}
The development process is organized in feature, integration, and release branches. The CI infrastructure supports the configuration of automated pre- and post-submit builds for each branch. In addition, each build can be configured for three different system architectures and multiple build profiles varying in compiler optimization levels and debug symbols. We incrementally activated the memory refinement approach based on the regression prediction model for all branches, build types, and system architectures while monitoring that the reduced memory assignments did not negatively affect build turnaround times or OOM failures.

\begin{figure}[htbp]
\centerline{\includegraphics[scale=.6]{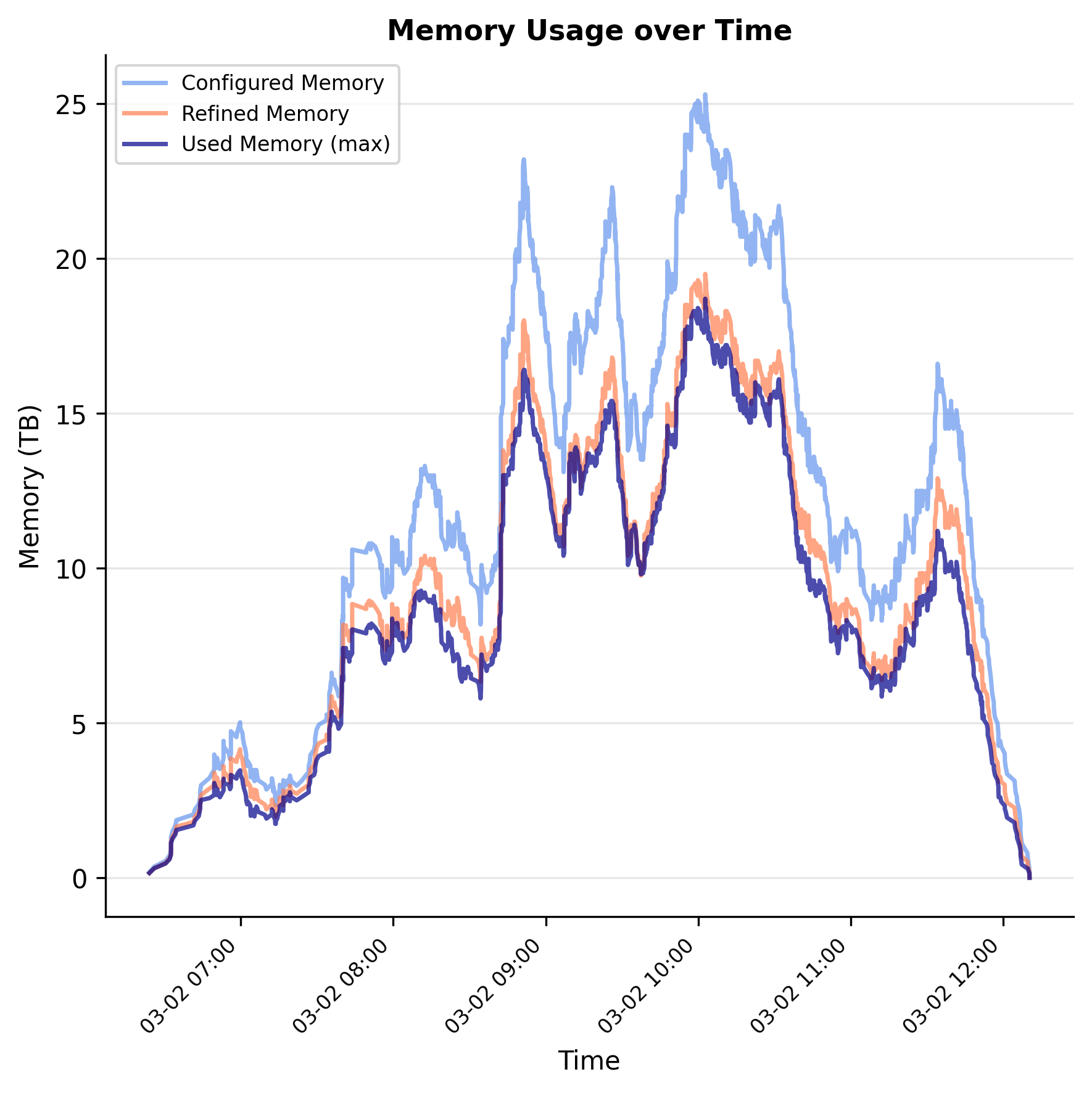}}
\caption{Aggregated cluster memory usage over a 6h period.}
\label{AggClusterMem}
\end{figure}

Figure~\ref{AggClusterMem} highlights the accuracy of our prediction model during a randomly chosen 6-hour interval. The chart compares the initial memory requirement configuration, the model-based refinement, and the measured system memory during C++ build job execution. The results indicate a significant reduction in build memory requirements, leading to better resource utilization on the task execution cluster. In addition, the memory measurements closely track the refined memory requirements, highlighting the accuracy of the regression model. 

\begin{figure}[htbp]
\centerline{\includegraphics[scale=.6]{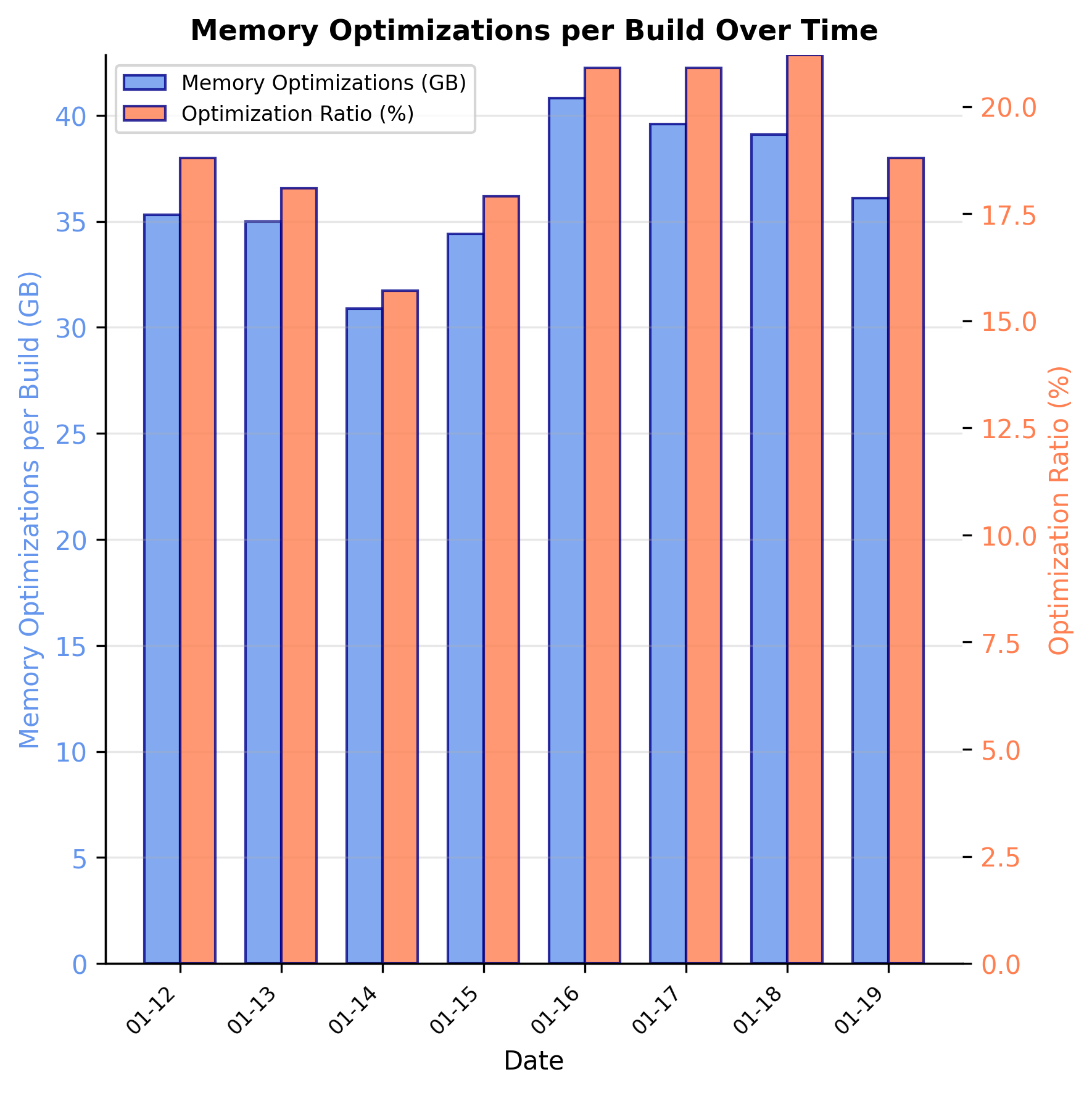}}
\caption{Average optimizations per day per build over one week.}
\label{AverageSavings}
\end{figure}

The average daily memory savings per build are shown in Figure~\ref{AverageSavings}. During the selected example week, we reached an average optimization of $\approx 36.4$GB per build. This result closely matches the experimental estimate of 37.2GB. In addition, the chart shows peak optimization ratios of more than 20\%.

\begin{figure}[htbp]
\centerline{\includegraphics[scale=.6]{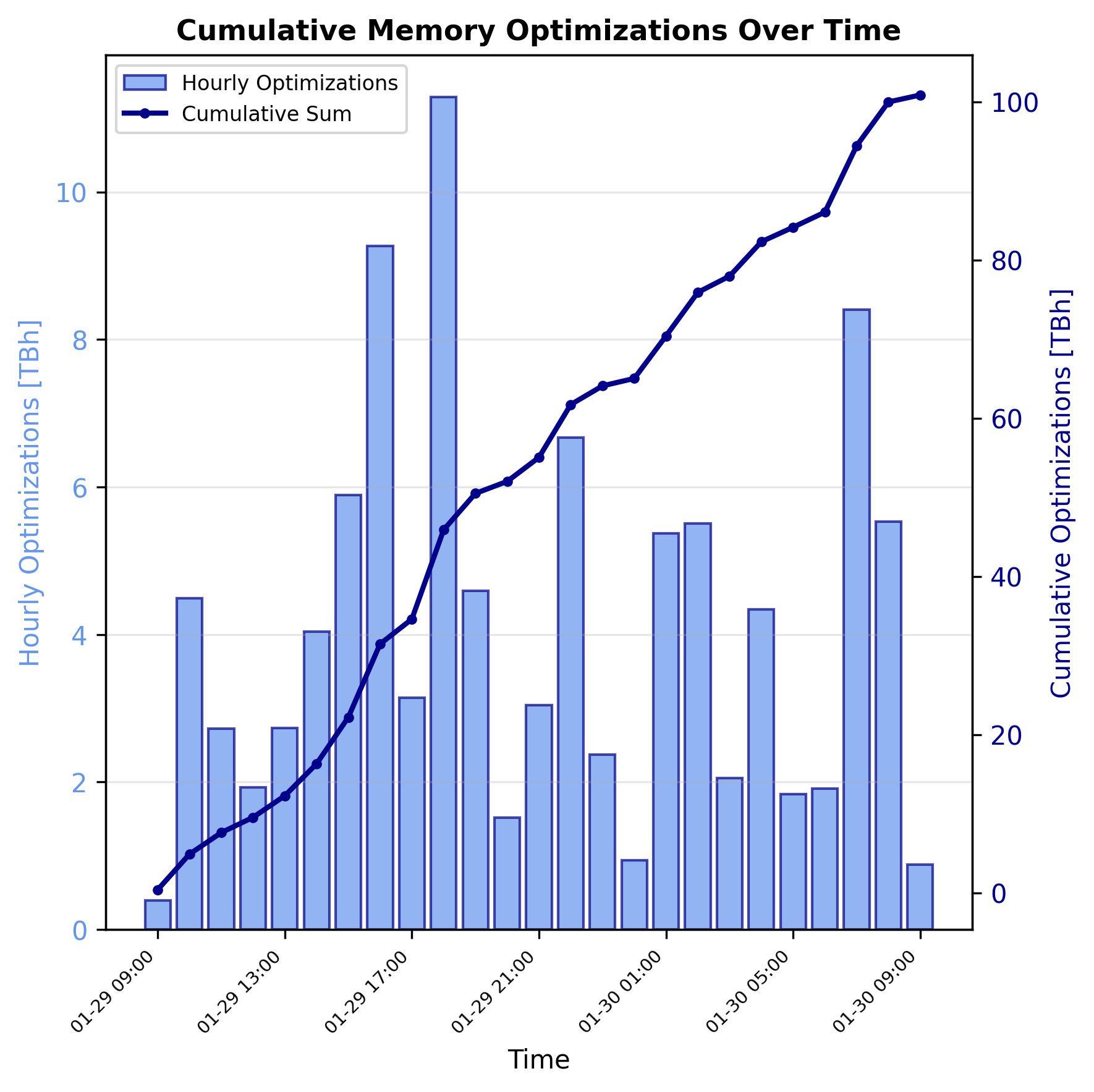}}
\caption{Timeline plot showing optimization instances and accumulated optimizations over 24 hours.}
\label{CumulativeWorth}
\end{figure}

In cloud environments, resource costs are often charged based on the time interval for which the resources are in use. Figure~\ref{CumulativeWorth} shows the system memory optimizations weighted by their duration. For example, a build job requiring 40GB less system memory over an execution time of 1 hour yields a memory resource optimization of 40GBh. The chart indicates a cumulative optimization of $\approx 100$TBh during a randomly chosen 24-hour period, which sums up to more than 2.8PBh per month. The optimization potential fluctuates over the course of the day depending on developer activity and incoming build requests.

\section{Related Work} \label{relatedWorkSection}

Memory prediction has been studied extensively in many areas such as cluster resource management, workflow execution, or scientific computing.
In the following, we describe existing work and compare it to our own method.

\subsection{Application Memory Prediction}

In this section, we cover publications that use a single prediction method for predicting application memory consumption.

Lehmann et al.~\cite{lehmann2024ponder} present Ponder, a tool that predicts the memory usage of scientific workflow applications.
During runtime, Ponder checks for a linear relationship between the input data and memory consumption.
Depending on this, a rule-based algorithm decides whether to apply a linear regression to predict the memory or to use the highest observed historical memory value.
Further, the authors implement a static offset to avoid failing executions due to under-allocations.

Bader et al.~\cite{bader2024ks+,baderPredictDynamicMemoryRequ2023} propose a method that uses time-series monitoring data to predict memory consumption over time.
Their prediction method consists of two stages.
In the first stage, the runtime is predicted using a linear regression model.
Based on this prediction, segments are created.
In the second stage, a linear regression model predicts the memory consumption for each segment.

Bader~et~al.~\cite{bader2022leveraging} also present two reinforcement learning memory prediction approaches.
Their reinforcement learning agents aim to minimize the amount of unused memory.
At the same time, the reward function discourages out-of-memory failures.
The approaches are based on gradient bandits and Q-learning, where memory allocations are modeled as discrete actions.
The agents iteratively adjust the assigned memory.

Witt~et~al.~\cite{witt2019feedback} propose two approaches, a percentile and a linear regression predictor, both of which operate during workload execution and update their respective predictions.
The percentile predictor estimates a certain percentile, e.g., the P90 percentile, based on historical executions.
The linear regression model offsets its prediction by the difference between the expected and actual memory usage.

Da Silva et al.~\cite{da2013toward,da2015online} predict the memory consumption of scientific workflow applications.
To this end, their method continuously monitors executions and updates predictions during runtime.
For the prediction, a regression tree trained on historical data is used.
The regression tree considers the workflow, the applications, the parameter to predict (e.g., memory or CPU), and whether a correlation exists between the application input and the parameter.

Instead of directly predicting the memory consumption of a workload, Tovar~et~al.~\cite{tovar2022dynamic} split the workload into multiple sub-tasks.
Their approach is applied in the domain of high-energy physics and requires workloads to be splittable.
While this requirement might not be fulfilled for many scenarios, such as predicting the memory consumption of a single application, other fields, such as tests bundled in test suites, could benefit from this approach.

\subsection{Ensemble Memory Prediction}

Some memory prediction approaches leverage multiple machine learning methods as an ensemble.
In this section, we take a closer look at such approaches.

Rodrigues~et~al.~\cite{rodrigues2016helping} present a memory prediction tool to help users set their HPC job memory requirements.
Their tool encompasses four machine learning methods: Support Vector Machines, Random Forests, Radial Basis Function networks, and k-Nearest Neighbors.
These methods are trained using online and offline data, while a sliding window emphasizes recently gathered data.
Furthermore, the approach formulates the problem as a classification task, using bin sizes in multiples of 512\,MiB.

Bader~et~al.~\cite{bader2024Sizey} introduce an online memory prediction interface that combines multiple machine learning models.
The authors use a resource allocation quality score that combines prediction accuracy and efficiency.
This score is used to weigh the models and produce a single prediction.
Lastly, the prediction is offset by a dynamically defined amount of memory that depends on historical observations for each task.

Tanash~et~al.~\cite{tanash2021ensemble} present a memory prediction approach for Slurm jobs that uses multiple machine learning models.
The initial model training is performed offline.
Based on evaluation scores, the best-performing model is selected for inference, while updates incorporating completed jobs are not performed.
The models are trained on a subset of job submission and execution features extracted from long-term Slurm accounting logs.

\subsection{Comparison with Own Work}

In contrast to the presented single-predictor methods, our approach uses an ensemble that combines LightGBM and XGBoost and selects the higher memory prediction, thereby avoiding costly under-predictions.
Similar to many existing approaches, we also employ a memory prediction offsetting technique to further reduce the risk of failing instances.
Furthermore, our approach runs in the SAP HANA CI/CD environment consisting of more than a thousand servers.
Thus, our implementation and experiments are not restricted to simulation and traces but provide insights into a large-scale production application.

\section{Conclusion and Future Work} \label{conclusionSection}

In this paper, we showed that intelligent memory prediction can substantially improve the efficiency of large-scale CI build environments. Our results provide three main insights: static memory classes can waste a considerable share of reserved capacity, conservative quantile ensembles provide a practical trade-off between savings and reliability, and a lightweight microservice-based orchestration layer is sufficient to operationalize model-driven resource refinement in production. For SAP HANA build workloads, this translates into substantial reductions in system memory usage and corresponding operational costs.

There are several opportunities for future extensions and improvements. 
In addition to build jobs, the CI workload of SAP HANA consists of thousands of unit, integration, and performance tests for each completed build.
As each test run has large system memory requirements based on the installation of the application, loading of test data, and execution of test logic, the optimization potential for system memory is quantitatively larger compared to the builds. 
Since test runtime often exceeds the build runtime, we expect the memory savings per unit of time to also be greater. 
We plan to evaluate and develop a memory prediction model for test workloads and integrate it into our new orchestration component.

The cluster management system allows the definition of multiple computational requirements for task execution.
In the current state, our modeling approach only considers the system memory requirements of tasks. 
For future work, we plan to consider multiple dimensions in our models and derive optimized predictions for both system memory and CPU resource requirements.


\section*{Acknowledgment}
We thank Bartosz Bogacz (SAP) for his initial work on the LinearSVC-based memory prediction model. 
We also thank Edgar Blumenthal, Marten Eckardt, Justus Krebs, Xemena Wysokinska, and Joel Witzke from the Technical University of Berlin for their collaboration on the development and evaluation of memory prediction models using offline execution data.

\bibliographystyle{plain} 
\bibliography{refs} 

\end{document}